\documentclass[aps,showpacs,nofootinbib,superscriptaddress]{revtex4}

\usepackage{graphicx}
\usepackage{bm}
\usepackage{amsmath}
\usepackage{amssymb}
\usepackage{hyperref}

\newcommand{\qslash}{\kern 0.2 em n\kern -0.50em /}
\newcommand{\nslash}{\kern 0.2 em n\kern -0.50em /}
\newcommand{\kslash}{\kern 0.2 em k\kern -0.45em /}
\newcommand{\lslash}{\kern 0.2 em l\kern -0.50em /}
\newcommand{\pslash}{\kern 0.2 em p\kern -0.50em /}
\newcommand{\Sslash}{\kern 0.2 em S\kern -0.50em /}
\newcommand{\Pslash}{\kern 0.2 em P\kern -0.50em /}
\newcommand{\Dslash}{\kern 0.2 em D\kern -0.65em /\kern 0.15em}

\newcommand{\tr}{\mbox{Tr}\,}

\usepackage{color}

\begin{document}

\title{The Transverse polarization of the $\Lambda$ hyperon from unpolarized quark fragmentation in the diquark model}
\author{Yongliang Yang}\affiliation{School of Physics, Southeast University, Nanjing
211189, China}
\author{Zhun Lu}\email{zhunlu@seu.edu.cn}\affiliation{School of Physics, Southeast University, Nanjing
211189, China}
\author{Ivan Schmidt}\email{ivan.schmidt@usm.cl}\affiliation{Departamento de F\'\i sica, Universidad T\'ecnica Federico Santa Mar\'\i a, and
Centro Cient\'ifico-Tecnol\'ogico de Valpara\'iso,
Casilla 110-V, Valpara\'\i so, Chile}

\begin{abstract}
We investigate the spin-dependent (naive) T-odd fragmentation function $D^\perp_{1T}$, which can provide an explanation on the transverse polarization of the $\Lambda^0$ hyperon produced in an unpolarized process. We calculate $D^\perp_{1T}$ for light flavors in the spectator diquark model, with a Gaussian form factor at the hyperon-quark-diquark vertex. We include in the calculation both the scalar diquark and axial-vector diquark spectators. We determine the values of the model parameters by fitting the unpolarized fragmentation function $D^\Lambda_1$ to the DSV parametrization for $D^\Lambda_1$. In addition, we compute the longitudinal polarization fragmentation function $G^\Lambda_1$ and compare it with the known parametrization of $G^\Lambda_1$. We also estimate the transverse polarizations of $\Lambda$ production, in both semi-inclusive deep inelastic scattering and single inclusive $e^+e^-$ annihilation.
\end{abstract}


\maketitle

\section{introduction}

The production of a polarized $\Lambda$ hyperon from unpolarized $pp$ collisions has been observed~\cite{Lesnik:1975my,Bunce:1976yb}. This  has become a long-standing challenge in high energy physics, since
it contradicts  the traditional theoretical expectation that single spin asymmetries in high energy scattering are forbidden at the partonic level and that the averaged polarization of the $\Lambda$ hyperon should be zero~\cite{Dharmaratna:1996xd}.
The production of the transversely polarized $\Lambda$ hyperon can therefore serve not only as a useful tool to study its spin structure~\cite{Ma:2001rm}, but also can provide further information on the
non-perturbative hadronization mechanism~\cite{Jaffe:1996wp,Burkardt:1993zh,Kanazawa:2015jxa,Boer:2010ya,Anselmino:2001js}.
Generally, the measurement of the $\Lambda$ polarization is quite difficult since its spin distribution is not directly accessible.
The self-analyzing properties of $\Lambda$ and a large angular distribution of the decay products (proton or pion) in the $\Lambda$ rest-frame~\cite{deFlorian:1997zj} afford a way to extract the polarization information of the $\Lambda$ hyperon.

While in the past a lot of experimental data and theoretical analyses provided us with information about the fragmentation functions for pion and kaon mesons, our knowledge of the $\Lambda$ fragmentation functions, particularly its polarized fragmentation function, is more limited.
This is so, in spite of the fact that the polarizations of the $\Lambda$ hyperon observed in $pp\rightarrow \Lambda\,X$ and $pp\rightarrow \Lambda^\uparrow (\text{jet})\, \text{jet}\,X$ reactions~\cite{Heller:1978ty,Heller:1983ia,Ramberg:1994tk,Smith:1986fz,Lundberg:1989hw,Pondrom:1985aw} have attracted theoretical studies and phenomenological analyses aiming at understanding the fragmentation mechanism behind the $\Lambda$ polarization~\cite{Boer:2007nh,Anselmino:2000vs,Collins:1992kk,Dong:2004qs,Sivers:1989cc,Felix:1999tf}.
A class of the so-called time-reversal-odd (T-odd) fragmentation functions has been the main focus of these efforts.
In particular, a leading-twist polarized fragmentation function, analogous to the Sivers function $f^\perp_{1T}$, denoted by $D^\perp_{1T}$, has been introduced in Refs.~\cite{Anselmino:2000vs,Mulders:1995dh}.
As a transverse momentum dependent (TMD) fragmentation function, $D^\perp_{1T}$ describes the fragmentation of an unpolarized quark to a transversely polarized hadron; and it may play an important role in the ¡°spontaneous¡± polarization, such as in: $q\rightarrow \Lambda^{\uparrow}X$~\cite{Boer:2009uc}.
Thus, a non-vanishing $D^{\perp \Lambda/q}_{1T}$ could help to illustrate the spin structure of the $\Lambda$ hyperon.
However, the single inclusive $e^+\,e^-$ annihilation (SIA) experiment performed by OPAL at LEP has not observed a significant signal on the transverse polarization of the $\Lambda$ hyperon~\cite{Ackerstaff:1997nh}.
As an alternative to SIA, the processes $e^+e^-\rightarrow \Lambda^\uparrow+h+X$ and the semi-inclusive deep inelastic scattering (SIDIS) $\ell\,p\rightarrow\ell'+\Lambda^\uparrow+X$ have been suggested~\cite{Yang:2016qsf,Boer:1997mf} to study the $\Lambda$ polarization, where $D^\perp_{1T}$ contribute to the cross section as well as to spin asymmetries.
One important result that validates these approaches is that the universality of fragmentation functions has been tested for different processes in Refs.~\cite{Meissner:2008yf,Gamberg:2008yt,Yuan:2007nd,Collins:2004nx,Metz:2002iz,Boer:2010ya}.
The single-inclusive $e^+\,e^-$ annihilation (SIA)~\cite{Acciarri:1997it,Alexander:1996qj} is similar to both the $pp$ collision and the SIDIS, and can play a similarly fundamental role in the measurement of the polarized $\Lambda$ production~\cite{Aid:1996ui,Arneodo:1984nf,Adams:1993qt}.
Recently, the Belle Collaboration presented the first observation of a nonzero transverse polarization of $\Lambda$ production in the inclusive
process $e^+e^-\rightarrow \Lambda(\bar{\Lambda})+X$ and $e^+e^-\rightarrow \Lambda(\bar{\Lambda})+K^\pm(\pi^\pm)+X$~\cite{Abdesselam:2016nym}.

Since the experimental information on $D^\perp_{1T}$ of the $\Lambda$ hyperon still remains unknown, model calculations will provide an approach to acquire knowledge of this quantity.
In this work we will calculate $D^\perp_{1T}$ of the $\Lambda$ hyperon for light flavors using a spectator model~\cite{Nzar:1995wb,Jakob:1997wg}, and study its contribution to the transverse polarization of the $\Lambda$ hyperon in SIDIS and SIA.
The spectator model has been applied to calculate the Collins fragmentation function of pion and kaon mesons~\cite{Bacchetta:2007wc},
as well as the twist-3 collinear fragmentation function of pion meson~\cite{Lu:2015wja,Yang:2016qsf}. In these cases the spectator system has been taken to be a quark.
The T-even fragmentation functions of the $\Lambda$ hyperon have also been calculated by the spectator model, in which case the spectator system is a diquark.
In our calculation, we will consider both the scalar diquark and the vector diquark, in order to obtain the flavor content of $\Lambda$ fragmentation functions.


The remaining content of this paper is organized as follows.
In Sec.~II we calculate the unpolarized fragmentation function $D^\Lambda_1$, as well as the longitudinally polarized fragmentation function $G^\Lambda_1$ for light flavors, using the diquark model.
The flavor decomposition of the fragmentation functions is realized by the SU(6) spin-flavor symmetry of the octet baryons.
We apply the values of the parameters which coincide with the DSV parametrization for $D^\Lambda_1$.
In Sec.~III, we use the same model and parameters to compute the T-odd fragmentation function $D^\perp_{1T}$ for up, down and strange quarks, considering the gluon scattering effect.
We then present numerical results of the transverse polarization of the $\Lambda$ hyperon in SIDIS and SIA.
Finally, we summarize our results and give conclusions in Sec.IV.

\section{Model calculation of unpolarized and longitudinally polarized $\Lambda$ fragmentation functions}

\begin{figure}
  \includegraphics[width=0.4\columnwidth]{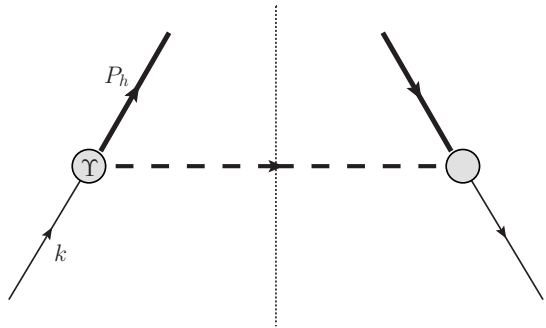}
 \caption {Lowest order diagram describing the fragmentation of a quark to a hyperon. The dashed line indicates a scalar diquark or an axial-vector diquark and the thick line is a $\Lambda$ hyperon}
 \label{tree}
\end{figure}

The unpolarized TMD fragmentation function $D^\Lambda_1(z,k_T)$ may be obtained from the following trace~\cite{Jakob:1997wg,Barone:2001sp}
 \begin{align}\label{Trace1}
 D^\Lambda_1(z,k_T)={1\over 4}\tr[(\Delta(z,k_T;S_{\Lambda})+\Delta(z,k_T;-S_{\Lambda})) \gamma^-] \,,
 \end{align}
where $\Delta(z,k_T;S_{\Lambda})$ is the TMD quark-quark correlation function~\cite{Bacchetta:2008af,Bacchetta:2006tn}
 \begin{align}\label{eq:delta1}
\Delta(z,k_T;S_\Lambda)&={1\over 2z}\int{dk^+\Delta(k,P_\Lambda;S_\Lambda)}\notag\\
&\equiv \sum_X\int{d\xi^+d^2\bm{\xi_T}\over 2z(2\pi)^3} e^{ik\cdot\,\xi}\langle 0|\, {\cal U}^{n^+}_{(+{\infty},\xi)}
\,\psi(\xi)|P_\Lambda,S_\Lambda; X\rangle\langle P_\Lambda,S_\Lambda; X|\bar{\psi}(0)\,
{\cal U}^{n^+}_{(0,+{\infty})}
|0\rangle \bigg|_{\xi^-=0}\,.
\end{align}
The Wilson line ${\cal U}$ is used to ensure gauge invariance of the operator~\cite{Collins:1981uw}, and
a detailed discussion on the structure of the Wilson line has been given in Ref.~\cite{Bacchetta:2007wc}.
The lowest-order diagram depicting the fragmentation of a quark into a $\Lambda$ hyperon in the spectator model is shown in Fig.~\ref{tree}, where
the final state $|P_\Lambda,S_\Lambda; X\rangle$ describes the outgoing $\Lambda$ hyperon and the intermediate unobserved states.
In this paper we perform the calculation in a diquark model~\cite{Nzar:1995wb,Jakob:1997wg}, which includes both the spin $0$ (scalar diquark) and spin $1$ (axial-vector diquark) spectator systems~\cite{Bacchetta:2008af,Yang:2002gh}.
The quark fragmenting (taking $u$ quark as an example) can be modeled as $u\rightarrow \Lambda(uds) + D(\bar{d}\bar{s})$, where $D$ denotes a diquark.
The matrix element which appears in the r.h.s. of Eq.~(\ref{eq:delta1}) has the following form
\begin{align}\label{eq:diquark}
  \langle\,P_\Lambda,S_\Lambda; X|\,\bar\psi(0)|0\rangle=
  \begin{cases}
  \bar{U}(P_\Lambda,S_\Lambda)\, {\Upsilon}_s\,\displaystyle{\frac{i}{\kslash-m_q}}&     \textrm{scalar diquark,} \\
 \bar{U}(P_\Lambda,S_\Lambda\,){\Upsilon}^{\mu}_v \,\displaystyle{\frac{i}{\kslash-m_q}}\,\varepsilon_{\mu}\, & \textrm{axial-vector diquark.}
  \end{cases}
\end{align}
Here $\Upsilon_{D}$ ($D=s$ or $v$) is the hyperon-quark-diquark vertex, and $\varepsilon_{\mu}$ is the polarization vector of the spin-1 axial-vector diquark.
In our work, the vertex structure is chosen as follows~\cite{Jakob:1997wg}
\begin{align}
  \Upsilon_s&=\bm{1}g_s\,,\notag\\
  \Upsilon^\mu_v &={g_v\over \sqrt{3}}\gamma_5(\gamma^\mu+{P_\Lambda^\mu\over M_\Lambda})
\end{align}
where $g_D$ ($D=s$ or $v$) is the suitable coupling for the hyperon-quark-diquark vertex.
For simplicity in this work we assume that $g_s$ and $g_v$ are the same: $g_s = g_v= g_D$.
Thus, the expression of the correlator in Fig.~\ref{tree} can be written as
\begin{align}\label{eq:dfey}
 \Delta(z,k_T;S_\Lambda)&=\frac{g_D^2}{4(2\pi)^3}{(\kslash +m_q)(\Pslash_\Lambda+ M_\Lambda)(1+a_D\gamma_5\Sslash_\Lambda)(\kslash +m_q)\over(1-z)P_\Lambda^-(k^2-m_q^2)^2}\,,
\end{align}
 with
 \begin{align}\label{eq:k2}
k^2={z\over (1-z)}\bm{k}^2_T+{m_D^2\over (1-z)}+{M_\Lambda^2\over z}\,,
\end{align}
and $k^-={P_\Lambda^-\over z}$.
The spin factor $a_D$ takes the values $a_s=1$ and $a_v=-{1\over 3}$, and
$m_q$, $m_D$ and $M_\Lambda$ represent the masses of the parent quark, the spectator diquark and the fragmenting $\Lambda$ hyperon, respectively.

Applying the diquark model, the unpolarized fragmentation function $D^\Lambda_1$ is derived from Eqs.~(\ref{Trace1}) and~(\ref{eq:dfey}):
\begin{align}\label{D1}
 D^{(s)}_1(z,z^2\bm k_T^2) = D^{(v)}_1(z,z^2\bm k_T^2) = {g_D^2\over 2(2\pi)^3}{(1-z)[z^2\bm{k}^2_T+(M_\Lambda+zm_q)^2]\over \,z^4(k_T^2+L^2)^2}\,,
\end{align}
where \begin{align}\label{L2}
L^2={1-z\over z^2}M_\Lambda^2+m_q^2+{m_D^2-m_q^2\over z}\,.
\end{align}
In Eq.~(\ref{D1}), $D^{(s)}_1(z,z^2\bm k_T^2)$ and $D^{(v)}_1(z,z^2\bm k_T^2)$ denotes the contributions to $D_1^\Lambda$ from the scalar diquark and the axial-vector diquark components, respectively, and the final results that we get for them turn out to be the same.
In order to obtain $D^{(v)}_1(z,z^2\bm k_T^2)$, we have used a summation for all polarizations states of the axial-vector diquark: $\sum_\lambda\varepsilon^{*(\lambda)}_\mu\varepsilon^{(\lambda)}_\nu=-g_{\mu\nu}+{P_{\Lambda\mu}\,P_{\Lambda\nu}\over M_\Lambda^2}$.

Assuming an SU(6) spin-flavor symmetry, the fragmentation functions of the $\Lambda$ hyperon for light flavors satisfy the following relations between the different quark flavors and diquark types~\cite{Hwang:2016ikf,VanRoyen:1967nq,Jakob:1993th}
 \begin{align}\label{relation}
D^{\textrm{u}\rightarrow \Lambda} =\,D^{\textrm{d}\rightarrow \Lambda} ={1\over 4}D^{(s)}+{3\over 4}D^{(v)}\,,~~D^{\textrm{s}\rightarrow \Lambda}=D^{(s)}\,,
\end{align}
where u, d and s denote the up, down and strange quarks, respectively.
In this study we assume that the relation (\ref{relation}) holds for all fragmentation functions.

Neglecting the mass differences between the up, down and strange quarks, the relation in Eq. (\ref{relation}) and the result in Eq.~(\ref{D1}) imply that the light quarks fragment equally to $\Lambda$ for the unpolarized fragmentation function $D^\Lambda_1$, $i.e.$
\begin{align}
D^{u\rightarrow\Lambda}_{1}=D^{d\rightarrow\Lambda}_{1}=D^{s\rightarrow\Lambda}_{1}\equiv D_1^\Lambda\,.
\end{align}
This result is consistent with the DSV parametrization of $D^{\Lambda}_{1}$ for light flavors presented in Ref.~\cite{deFlorian:1997zj}, in which the $e^+e^-\rightarrow \Lambda\,X$ data were applied to perform the corresponding fit.

One can perform the integration over the transverse momentum of the produced hadron $\bm P_{T}=-z\bm{k}_T$ w.r.t. the quark direction, to obtain the integrated unpolarized fragmentation function $D^\Lambda_1(z)$:
 \begin{align}
 D^\Lambda_1(z)= \int d^2\bm P_T D_1^\Lambda(z, \bm P_T^2)=\pi\,z^2\int^\infty_0{dk_T^2\,D_1^\Lambda(z,z^2\bm k_T^2)}\,,
 \end{align}
which is divergent from large values of $k_T$, when a point-like hyperon-quark-diquark coupling is considered.
In Refs.~\cite{Gamberg:2003eg,Bacchetta:2007wc,Amrath:2005gv}, two different approaches to regularize this divergence are presented.
One of them is to set an upper limit for $k_T$, while the other is to choose a $k^2$ dependent Gaussian form factor for the hyperon-quark-diquark coupling:
\begin{align}\label{factor}
  g_D~\mapsto {g_D\over z}\,e^{-{k^2\over \Lambda^2}}\,,
\end{align}
where $\Lambda^2$ has the general form $\Lambda^2= \lambda^2z^\alpha(1-z)^\beta$.
In this work
we choose a Gaussian form factor, since with this choice the unpolarized fragmentation function can be reproduced reasonably well.
Due to the relation between $k^2$ and $\bm{k}_T^2$, presented in Eq.~(\ref{eq:k2}), the divergence arising from the large $k_T$ region can be effectively cut off.
Thus, the analytic result for $D^\Lambda_1(z)$ is
 \begin{align}\label{ID1}
  D^{\Lambda}_1(z)=&{g_s^2\over 4(2\pi)^2}{e^{-{2m_q^2\over\Lambda^2}}\over z^4 L^2}\bigg{\{}z(1-z)((m_q+ M_\Lambda)^2 - m_D^2)\exp\biggl({-2zL^2\over (1-z)\Lambda^2}\biggr) \notag\\
  &+\bigl((1-z)\Lambda^2-2((m_q+M_\Lambda)^2-m_D^2)\bigr){z^2L^2\over\Lambda^2}\Gamma\biggl(0,{2zL^2\over(1-z)\Lambda^2}\biggr) \bigg{\}}\,,
 \end{align}
 where the incomplete gamma function has the form
 \begin{align}\label{gammaf}
  \Gamma(0,z)~\equiv\,\int_z^\infty{e^{-t}\over t}dt\,.
 \end{align}
The parameters of the model are $\lambda, \alpha, \beta$, together with the masses of the spectator diquark $m_D$ and the parent quark $m_q$.

In order to get numerical result we choose the constituent quark mass as $m_q= 0.36\,\text{GeV}$ for up, down and strange quarks, and the $\Lambda$ hyperon mass as $1.116\,\text{GeV}$.
For the values of the other parameters, we fit our model result to the leading order (LO)  DSV parametrization for $D^\Lambda_1$ at the initial scale $\mu^2_{LO}=0.23~\textrm{GeV}^2$.
We note that $D^\Lambda_1$ given in Ref.~\cite{deFlorian:1997zj} is for the fragmentation of quarks to $\Lambda^0 + \bar\Lambda^0$.
On the other hand, in the diquark model one can only calculate the valence quark contribution (favored) to the $\Lambda$ hyperon, such as $\text{u}\rightarrow \Lambda^0$ or $\bar{\text{u}}\rightarrow \bar{\Lambda}^0$; while the sea quark contribution (unfavored, e.g.,$\text{u}\rightarrow \bar{\Lambda}^0$) is zero. In order to mimic the unfavored fragmentation function we assume that it is proportional to the favored fragmentation function, and therefore the unfavored fragmentation function can be also included in the model by adjusting the coupling $g_D$.

\begin{figure}
  \centering
  \includegraphics[width=0.48\columnwidth]{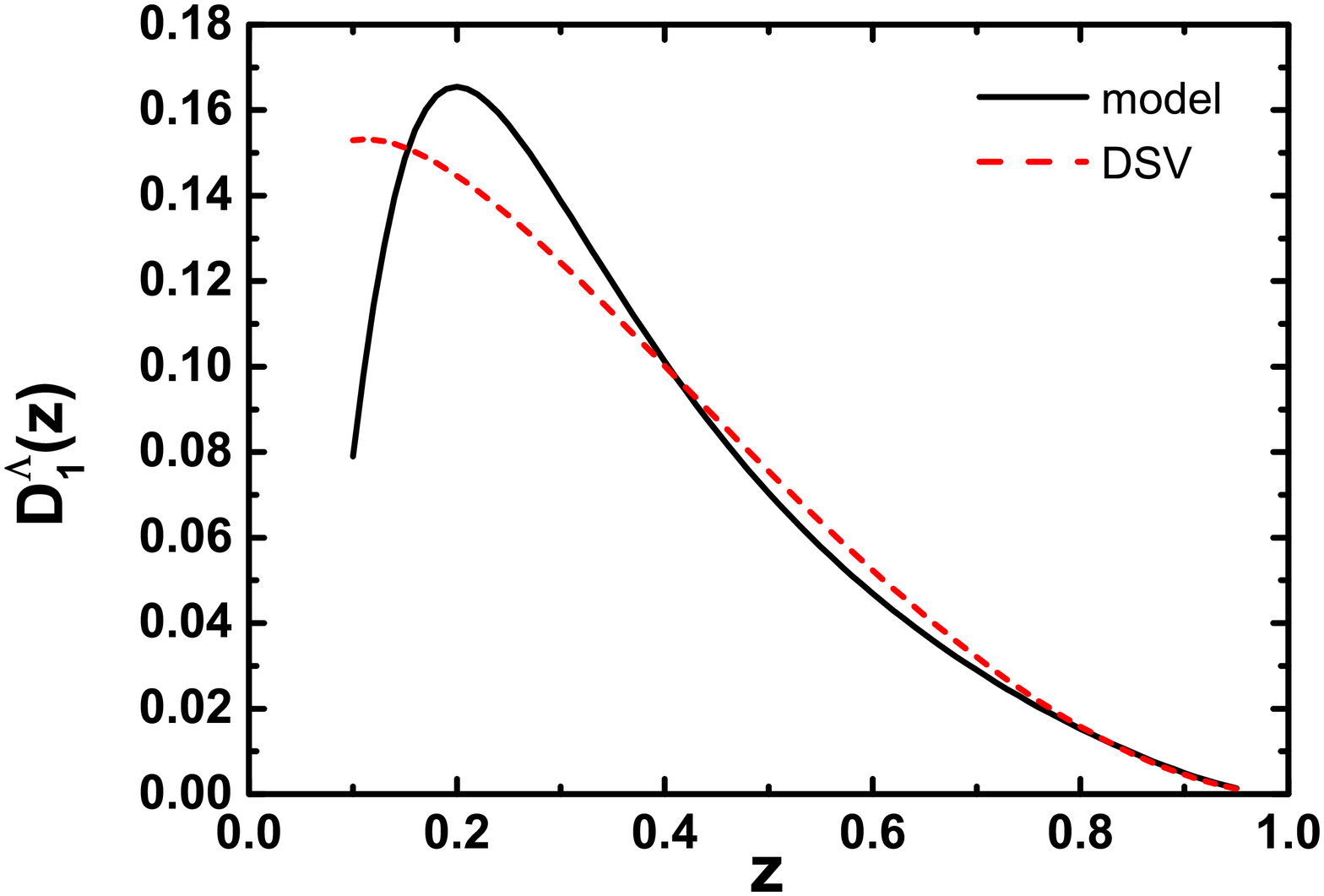}
  \includegraphics[width=0.48\columnwidth]{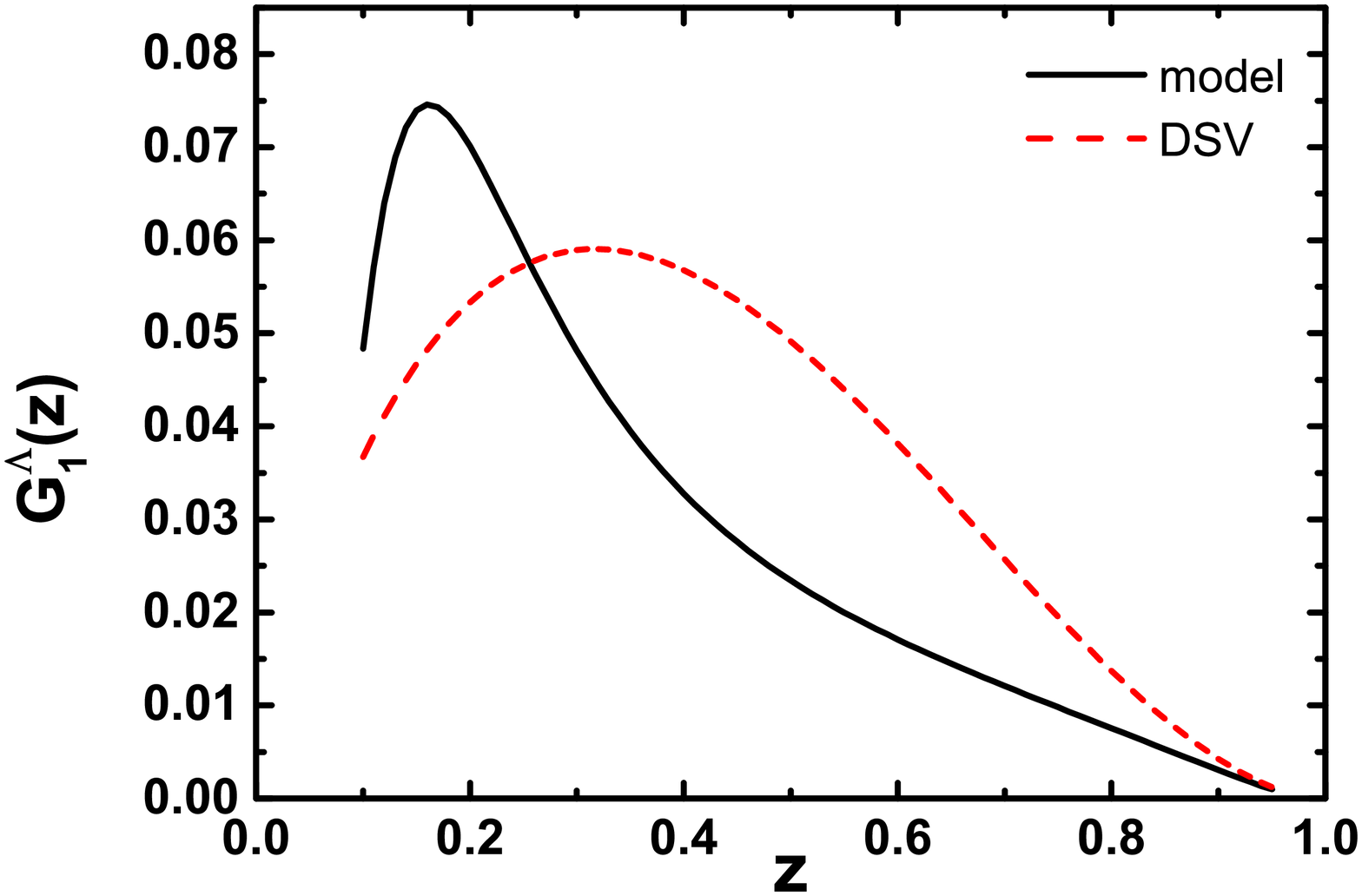}
  \caption{Unpolarized fragmentation function $D^\Lambda_1(z)$ vs $z$ (left panel), the polarized fragmentation functions $G^{\Lambda}_1(z)$ vs. $z$( right panel).}
  \label{fig1}
\end{figure}

The values of the parameters $\alpha, \beta$ are fixed in the fit.
The fitted results are
    \begin{align}\label{paramters}
      g_D=1.983,~~~ m_D=0.745\textrm{GeV},~~~ \lambda=5.967\textrm{GeV},~~~ \alpha=0.5(\textrm{fixed}),~~~ \beta=0(\textrm{fixed}).
     \end{align}
In the left panel of Fig.~\ref{fig1}, we plot our model calculation of the unpolarized fragmentation function $D^{q\rightarrow \Lambda}_1(z)$ (solid line), using the parameters presented in Eq.~(\ref{paramters}).
The parametrization of the DSV~\cite{deFlorian:1997zj} is also shown for comparison (dashed line).

Using the values of the parameters fitted from $D_1^\Lambda(z)$, we also calculate the light flavor fragmentation functions for the longitudinally polarized $\Lambda$ hyperon, denoted by $G_{1}^\Lambda(z)$, as a cross check of our calculation.
The polarized TMD fragmentation function $G_{1L}$ can be obtained from the following trace:
\begin{align}\label{Trace2}
{1\over 4}\tr[(\Delta(z,k_T;S_{\Lambda})-\Delta(z,k_T;-S_{\Lambda})) \gamma^-\gamma_5] = S_{\Lambda{L}}\,G_{1L}- {k_{T}\cdot\,S_{\Lambda{T}}\over M_\Lambda}G_{1T}\,,
\end{align}
where the spin vector of the $\Lambda$ hyperon is decomposed as
\begin{align}\label{spin}
S_{\Lambda}^\mu = S_{\Lambda\,L}{(P_\Lambda\cdot n_+)n_-^\mu\,- (P_\Lambda\cdot n_-)n_+^\mu\over M_\Lambda}+ S_{\Lambda\,T}^\mu\,.
\end{align}
Applying Eqs.~(\ref{eq:dfey}), (\ref{Trace2}) and (\ref{spin}), we arrive at the following expression for $G^{(D)}_{1L}(z,k_T^2)$ in the diquark model:
\begin{align}\label{G1}
G^{(D)}_{1L}(z,k_T^2)&=-a_D{g_D^2\over 2(2\pi)^3}{(1-z)[z^2k_T^2-(z\,m_D+M_\Lambda)^2]\over z^4(k_T^2+L^2)^2}\,.
\end{align}

With the relation between different quark flavors and diquark types for the polarized fragmentation functions given in Eq.~(\ref{relation}), we obtain the light flavor fragmentation function $G_{1L}^\Lambda$ as follows
\begin{align}\label{relation1}
G_{1L}^{\textrm{u}\rightarrow \Lambda}(z,k_T^2)= 0\,,~~~ G_{1L}^{\textrm{d}\rightarrow \Lambda} (z,k_T^2)=0\,,~~~G_{1L}^{\textrm{s}\rightarrow \Lambda}(z,k_T^2)=G_{1L}^{(s)}(z,k_T^2).
\end{align}
We find that although both the scalar diquark and axial-vector diquark components contribute to $G_{1L}^\Lambda$, they cancel exactly and yield vanishing polarized fragmentation functions for the up and down quarks.
The strange quark fragmentation function $G_{1L}^{s\rightarrow \Lambda}$ survives and it only receives contribution from the scalar diquark.
This result is consistent with the scenario 1 parametrization for the polarized fragmentation function $G_{1}(z)$ in Ref.~\cite{deFlorian:1997zj}, where only the strange quark contribution to polarized $\Lambda$ production is considered.

The integrated fragmentation function $G^\Lambda_1(z)$ is defined as
\begin{align}
 G^\Lambda_1(z)=\pi\,z^2\int^\infty_0{dk_T^2\,G^{\textrm{s}\to\Lambda}_{1L}(z,z^2\bm k_T^2)}\,.
 \end{align}
Here we take the same choice for the form factor as in the calculation of $D^\Lambda_1$, which leads to the following result
 \begin{align}\label{IG1}
  G^{\Lambda}_1(z)=&{g_s^2\over 4(2\pi)^2}{e^{-{2m_q^2\over\Lambda^2}}\over z^4 L^2}\bigg{\{}(1-z)\bigr[M_\Lambda^2(2-z) +2z\,m_qM_\Lambda +z(m_D^2 + m_q^2(2z-1))\bigr]\exp\biggl({-2zL^2\over (1-z)\Lambda^2}\biggr) \notag\\
  &-z[2 M_\Lambda^2(2-z)+4z\,m_qM_\Lambda+z\bigr((1-z)\Lambda^2 +2(m_D^2 + m_q^2(2z-1))\bigr)]{L^2\over\Lambda^2}\Gamma\biggl(0,{2zL^2\over(1-z)\Lambda^2}\biggr) \bigg{\}}\,.
 \end{align}

In the right panel of Fig.~\ref{fig1}, we plot our numerical result for $G^{\textrm{s}\rightarrow\Lambda}_1(z)$ vs $z$ (solid line) and compare it with the parametrization for $G^{\Lambda}_1(z)$ within scenario 1 (dashed line) in Ref.~\cite{deFlorian:1997zj} .
We find that it qualitatively agrees with the DSV parametrization, although at the regime $z\ge 0.3$ the size of the model result is smaller than that of the parametrization, which might be explained by the fact that in the experiment part of the measured polarized $\Lambda$ is produced from the decay of heavier hyperons.

\section{Model calculation of the T-odd fragmentation function $D^\perp_{1T}$ }

In this section, we calculate the T-odd TMD fragmentation function $D^{\perp}_{1T}$, which describes the number density of a transversely polarized $\Lambda$ hyperon fragmented from an unpolarized quark~\cite{Anselmino:2001js,Bacchetta:2004jz}£º
\begin{align}
D_{\Lambda^\uparrow/q}(z,\bm{P}_T) -  D_{\Lambda^\downarrow/q}(z,\bm{P}_T)=
\Delta D_{\Lambda^\uparrow/q}(z,\bm{P}_T^2){(\hat{\bm{k}} \times \bm{P}_T)\cdot \bm{S}_\Lambda \over z M_\Lambda}\,,
\end{align}
where $\hat{\bm{k}}$ is the unit vector along the fragmenting quark, and $\Delta D_{\Lambda^\uparrow/q}$ is an alternative notation for $D^\perp_{1T}$ defined in Ref.~\cite{Anselmino:2001js}, which is related to $D^\perp_{1T}$ by
\begin{align}
\Delta D_{\Lambda^\uparrow/q}(z,\bm{k}_T^2)={|\bm{P}_{T}|\over zM_\Lambda}D^{\perp\,q}_{1T}(z,\bm{k}_T^2)={|\bm{k}_{T}|\over M_\Lambda}D^{\perp\,q}_{1T}(z,\bm{k}_T^2)\,.
\end{align}
Notice the appearance of the vectorial triple product, which indicates a (naive) T-odd expression, since it has two momenta and one spin vectors.

Following Ref.~\cite{Anselmino:2000vs}, $D^\perp_{1T}$ can be obtained from the following trace:
\begin{align}\label{eq:TraceD}
&{\epsilon_T^{\rho\sigma} k_{T\rho}\,S_{\Lambda\,T\sigma}\over M_\Lambda}D^\perp_{1T} (z,k_T)={1\over 4}\textrm{Tr}[(\Delta(z,k_T;S_{\Lambda\,T})-\Delta(z,k_T;-S_{\Lambda\,T})) \gamma^-]\,.
\end{align}
\begin{figure}
  \includegraphics[width=0.52\columnwidth]{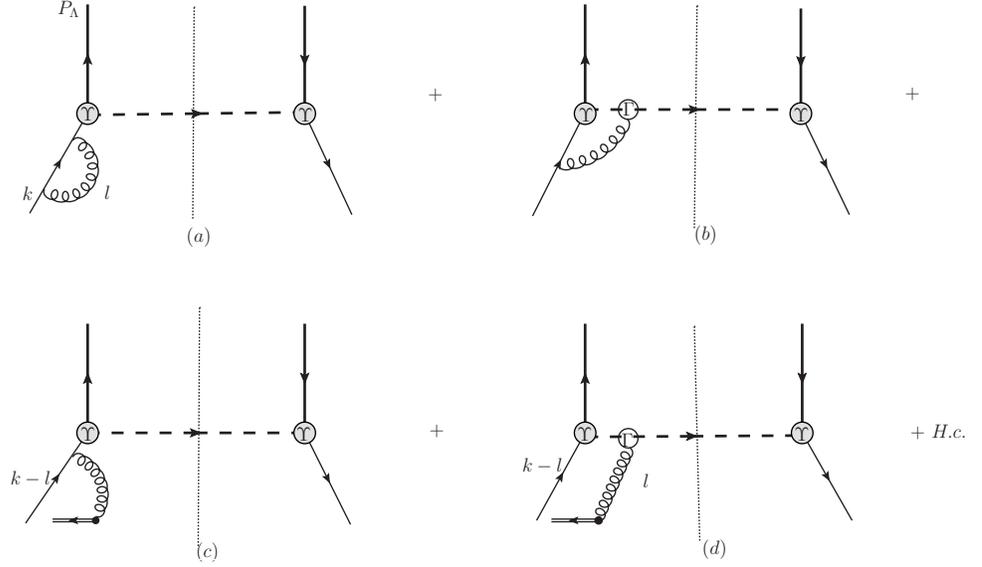}
 \caption {One loop corrections to the fragmentation of a quark into a $\Lambda$ hyperon in the spectator model. The double lines in (c) and (d) represent the eikonal lines. Here ``H.c." stands for the hermitian conjugations of these diagrams.}
 \label{loop}
\end{figure}
As is well-known, the tree-level calculation cannot provide a contribution to T-odd fragmentation functions, because of the lack of final or initial state interactions to produce imaginary phases in the scattering amplitude~\cite{Brodsky:2002cx,Brodsky:2002rv}.
The necessary nonzero contribution comes from loop corrections.
At one loop level, there are four diagrams that can generate imaginary phases, as shown in Fig.~\ref{loop}.
In Figs.~\ref{loop}(b)and~\ref{loop}(d), the notation $\Gamma$ is used to depict the gluon-diquark vertex, and we apply the following rules for the vertex between the gluon and the scalar diquark ($\Gamma_s$) and the axial vector diquark ($\Gamma_v$):
\begin{align}
\Gamma^{\rho,a}_s &=i\,g\,T^a\,(2k-2P_\Lambda-l)^\rho\,,\\
\Gamma_v^{\rho,\mu\nu,a} &=-i\,g\,T^a\,\big{[}(2k - 2P_\Lambda -l )^\rho g^{\mu\nu}-(k-P_\Lambda-l)^\nu g^{\rho\mu}-(k-P_\Lambda)^\mu g^{\rho\nu}\big{]}\,.\label{eq:factors}
\end{align}
Here, $T^a$ is the Gell-Mann matrix, and $g$ is the coupling constant of QCD.
Since the $\Lambda$ hyperon is colorless, it is expected that the spectator diquark should have the same color as that of the parent quark.
The Feynman rules for the eikonal line and vertex with gluon can be found in Refs.~\cite{Collins:1981uk,Collins:1981uw,Bacchetta:2007wc}.

Performing the integration over the loop momentum $l$, using the Cutkosky cutting rules, we first give the expression for $D^\perp_{1T}$, coming from the scalar diquark component
\begin{align}\label{eq:Dperps}
D^{\perp\,(s)}_{1T}(z,k_T^2) & =  {\alpha_s g_s^2C_F\over (2\pi)^4 }{e^{-2k^2\over \Lambda^2}\over z^2(1-z)}{1\over (k^2-m^2)}\left(D^{\perp\,(s)}_{1T(a)} (z,k_T^2) +D^{\perp\,(s)}_{1T(b)} (z,k_T^2) +D^{\perp\,(s)}_{1T(c)}(z,k_T^2)+D^{\perp\,(s)}_{1T(d)} (z,k_T^2)\right)\,,
\end{align}
where the four terms in the brackets correspond to the contributions from the four diagrams (plus their hermitian conjugates) in Fig.~\ref{loop}, respectively, and they read
\begin{align}
\begin{split}
D^{\perp\,(s)}_{1T(a)} (z,k_T^2) &= {m_q\,M_\Lambda\over (k^2-m_q^2)}(3-{m_q^2\over k^2})\,I_1\,,
\end{split}\displaybreak[0]\\
\begin{split}
D^{\perp\,(s)}_{1T(b)} (z,k_T^2) &= {M_\Lambda}\bigg{\{}m_q(2I_2-\mathcal{A})-M_\Lambda(\mathcal{B}-2I_2+2\mathcal{A})\bigg{\}}\,,
\end{split}\displaybreak[0]\\
\begin{split}
D^{\perp\,(s)}_{1T(c)} (z,k_T^2) &=0\,,
\end{split}\displaybreak[0]\\
\begin{split}
D^{\perp\,(s)}_{1T(d)} (z,k_T^2) &={M_\Lambda\over z}\bigg{\{}2(1-z)(m_q\mathcal{C}P_\Lambda^- -M_\Lambda\,\mathcal{D}P_\Lambda^-) -z(M_\Lambda\,\mathcal{B} -m_q\mathcal{A})\bigg{\}}\,.
\end{split}\displaybreak[0]
\end{align}
Here $\mathcal{A}$, $\mathcal{B}$, $\mathcal{C}$ and $\mathcal{D}$ are functions of $k^2$, $m_q$, $m_D$ and $M_\Lambda$,
\begin{align}
\mathcal{A}&={I_1\over \lambda(M_\Lambda,m_D)} \left(2k^2 \left(k^2 - m_D^2 - M_\Lambda^2\right) {I_{2}\over \pi}+\left(k^2+M_\Lambda^2 - m_D^2\right)\right)\,,\notag\\
\mathcal{B}&=-{2k^2 \over \lambda(M_\Lambda,m_D) } I_{1}\left (1+{k^2+m_D^2-M_\Lambda^2 \over \pi} I_{2}\right),\notag\\
\mathcal{C}P_\Lambda^- &={I_{34}\over 2k_T^2} +{1\over 2zk_T^2}\left(-zk^2 + \left(2-z\right) M_\Lambda^2 + zm_D^2 \right)I_2, \notag\\
\mathcal{D}P_\Lambda^- &={-I_{34}\over 2zk_T^2} -{1\over 2zk_T^2}\left(\left(1-2z\right)k^2 + M_\Lambda^2 - m_D^2 \right)I_2. \notag
\end{align}
The functions $I_{i}$ in the above equations are defined as
\begin{align}
I_{1} &=\int d^4l \delta(l^2) \delta((k-l)^2-m_q^2) ={\pi\over 2k^2}\left(k^2-m_q^2\right)\,, \\
I_{2} &= \int d^4l { \delta(l^2) \delta((k-l)^2-m_q^2)\over (k-P_\Lambda-l)^2-m_D^2}
={\pi\over 2\sqrt{\lambda(M_\Lambda,m_D)} }  \ln\left(1-{2\sqrt{ \lambda(M_\Lambda,m_D)}\over k^2-M_\Lambda^2+m_D^2 + \sqrt{ \lambda(M_\Lambda,m_D)}}\right)\,,\\
I_{34} &= \pi\ln{\sqrt{k^2}(1-z)\over m_D}\,,
\end{align}
with $\lambda(M_\Lambda,m_D)=(k^2-(M_\Lambda+m_D)^2)(k^2-(M_\Lambda-m_D)^2)$.

Note that when calculating the diagrams in Fig. \ref{loop}b and \ref{loop}d, we chose that the form factor $g_D$ depend only on the initial quark momentum $k$ instead of the loop momentum $l$.
This simplifies the integration over $l$, since the main effect of the form factor is to introduce to cut off to the high $k_T$ region. The same choice has also been used in Refs.~\cite{Bacchetta:2007wc,Lu:2015wja},

Similarly, using the gluon vertex given in Eq.~(\ref{eq:factors}), we can also calculate the expression for $D^\perp_{1T}$ from the axial vector diquark component
\begin{align}\label{eq:Dperpv}
D^{\perp\,(v)}_{1T}(z,k_T^2) & =   {2\alpha_s g_s^2C_F\over (2\pi)^4 }{e^{-2k^2\over \Lambda^2}\over z^2(1-z)}{1\over M_\Lambda (k^2-m_q^2)}\left(D^{\perp\,(v)}_{1T(v)} (z,k_T^2) +D^{\perp\,(v)}_{1T(b)} (z,k_T^2) +D^{\perp\,(v)}_{1T(c)}(z,k_T^2)+D^{\perp\,(v)}_{1T(d)} (z,k_T^2)\right)\,,
\end{align}
where the four terms in the r.h.s. of Eq.~(\ref{eq:Dperpv}) are given by:
\begin{align}
\begin{split}
D^{\perp\,(v)}_{1T(a)} (z,k_T^2) &={-m_q\,M_\Lambda\over (1-z)(k^2-m_q^2)}\left(1-{m_q^2\over 3k^2}\right)\,I_1\,,
\end{split}\displaybreak[0]\\
\begin{split}
D^{\perp\,(v)}_{1T(b)} (z,k_T^2) &={1\over 3(k^2-m_q^2)}\bigg{\{}2M_\Lambda[m_q(I_2-\mathcal{A})+M_\Lambda(\mathcal{A}-I_2-\mathcal{B})] +k\cdot P_\Lambda(4I_2-6\mathcal{A})\\
&-(\mathcal{A} k\cdot P_\Lambda+\mathcal{B} P_\Lambda^2)+{3\over 2}\left({k^2-m_q^2\over 2k^2}I_1+(k^2-m_q^2)\mathcal{A}\right)\bigg{\}}\,,
\end{split}\displaybreak[0]\\
\begin{split}
D^{\perp\,(v)}_{1T(c)} (z,k_T^2) &=0\,,
\end{split}\displaybreak[0]\\
\begin{split}
D^{\perp\,(v)}_{1T(d)} (z,k_T^2) &={-1\over 3M_\Lambda(k^2-m_q^2)}\bigg{\{}[M_\Lambda\bigr((k^2-m_q^2)\mathcal{C}P_\Lambda^-
+2M_\Lambda^2\mathcal{D}P_\Lambda^-\\
&-2m_qM_\Lambda\mathcal{C}P_\Lambda^-\bigr)+2k\cdot P_\Lambda(m_q\,\mathcal{C}P_\Lambda^- -M_\Lambda\,\mathcal{D}P_\Lambda^-)+z{m_q\over 2}I_1+{(k^2-m_q^2)\over 2}(M_\Lambda\mathcal{D}P_\Lambda^--m_q\mathcal{C}P_\Lambda^-)] \\
&-M_\Lambda(m_qM_\Lambda\,\mathcal{A}+2k\cdot P_\Lambda\mathcal{A}+M_\Lambda^2\mathcal{B})-{2M_\Lambda\over z}(m_q\,M_\Lambda\mathcal{C}P_\Lambda^-+k\cdot P_\Lambda\,\mathcal{C}P_\Lambda^-)\bigg{\}}\,.
\end{split}
\end{align}


As in Eq.~(\ref{relation}), the same relations should also hold for the fragmentation function $D^{\perp}_{1T}$:
\begin{align}\label{relation2}
D^{\perp\,u}_{1T} = D^{\perp\,d}_{1T} ={1\over 4}D^{\perp(s)}_{1T}+{3\over 4}D^{\perp(v)}_{1T}\,,~~D^{\perp\text{s}}_{1T}=D_{1T}^{\perp(s)}\,.
\end{align}
We apply the above equations to obtain $D^{\perp }_{1T}$ for light flavors and calculate the half $k_T$-moment of $D^{\perp}_{1T}$, which is defined as:
\begin{align}
D^{\perp(1/2)}_{1T}(z) = z^2\,\int{d\bm{k}^2_T}{|\bm{k}_T|\over 2M_\Lambda}D^{\perp}_{1T}(z,\bm{k}_T^2)\,.
\end{align}

As a leading-twist fragmentation function, the T-odd fragmentation function $D^\perp_{1T}$ should obey the following positivity bound~\cite{Bacchetta:1999kz}, which is an important theoretical constraint:
\begin{align}
{|k_T|\over M_\Lambda}D^{\perp}_{1T}(z,\bm{k}_T^2) \leq\,D_1(z,\bm{k}_T^2)\,.
\end{align}
Integration over $k^2_T$ gives an approximate expression for the positivity bound in terms of $D^{\perp(1/2)}_{1T}(z)$
\begin{align}
2D^{\perp(1/2)}_{1T}(z) \leq\,D_1(z)\,.
\end{align}

\begin{figure}
  \centering
  \includegraphics[width=0.48\columnwidth]{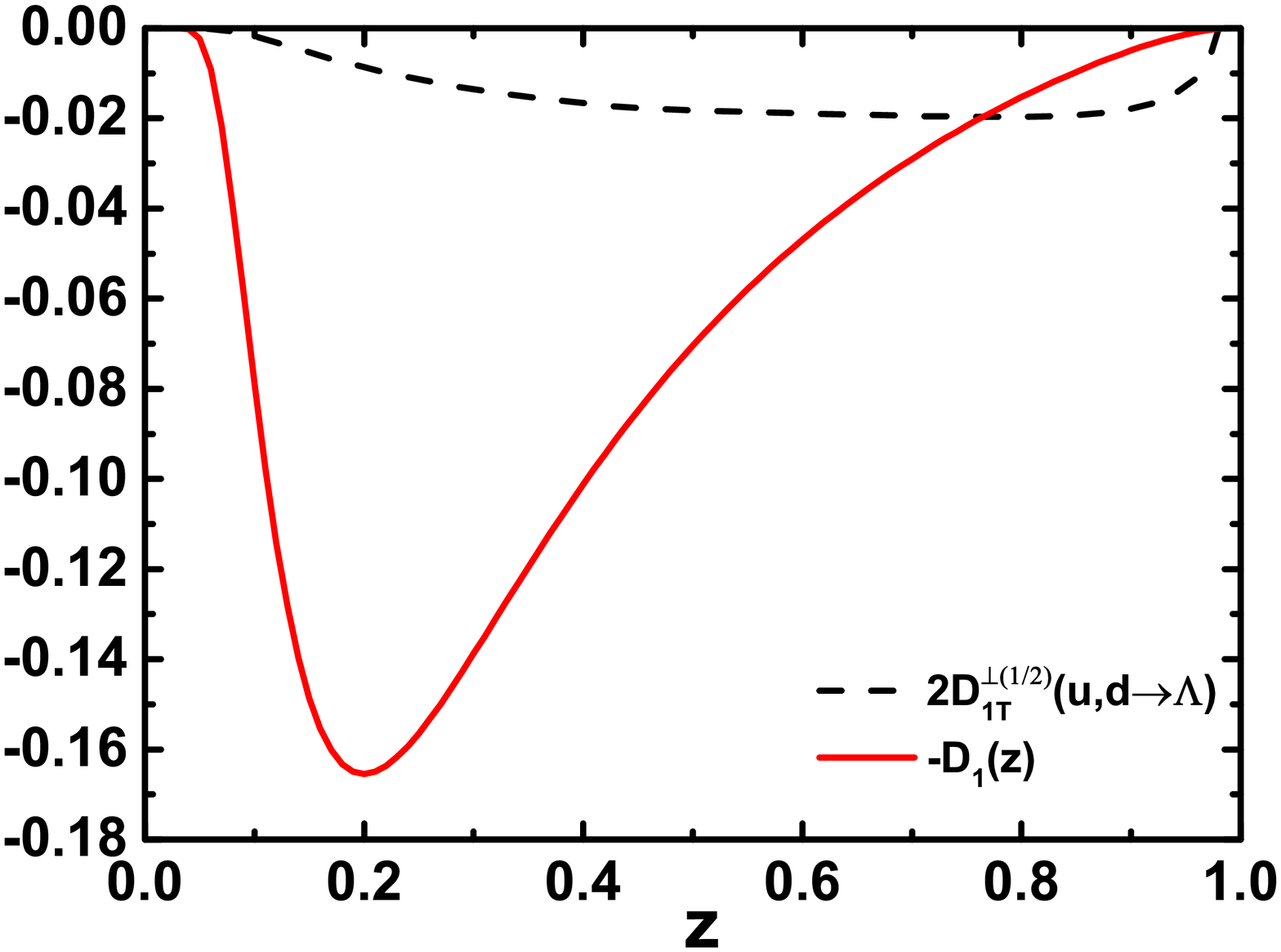}
  \includegraphics[width=0.48\columnwidth]{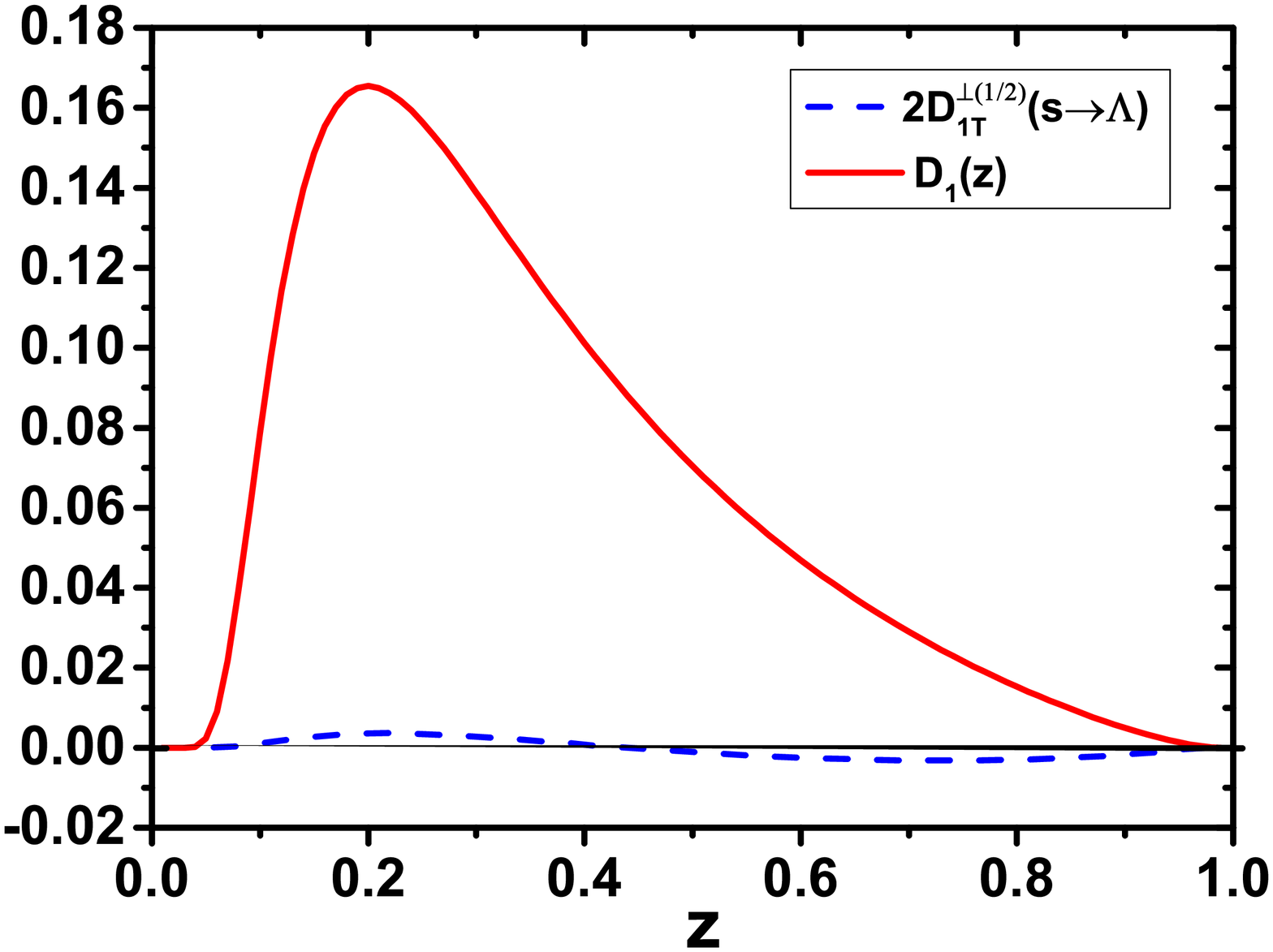}
  \caption{Left panel: the $D^{\perp(1/2)}_{1T}(z)$ (multiplied by 2) (dashed line) and $D_1(z)$ (multiplied by -1) (solid line) of the up and down quark. Right panel: the $D^{\perp(1/2)}_{1T}(z)$ (multiplied by 2) (dashed line) and $D_1(z)$ (solid line) of the strange quark.}
  \label{compare1}
\end{figure}

Using the parameters presented in Eq.~(\ref{paramters}), we calculate the half $k_T$-moment of the $\Lambda$ fragmentation function $D^{\perp}_{1T}$ for light flavors at the model scale $Q^2=0.23\,\text{GeV}^2$. In the calculation we choose the strong coupling constant at the model scale as $\alpha_s(\mu_0^2)=0.817$.
The numerical result of $D^{\perp(1/2)}_{1T}(z)$ (multiplied by a factor of 2) is plotted in Fig.~\ref{compare1}, in which the dashed line in the left panel shows the curves for the u and d quarks. The result for the s quark is shown in the right panel. The unpolarized fragmentation function $D_1(z)$ (solid lines) is also plotted as the positivity bound for comparison.
We find that the size of $D^{\perp(1/2)}_{1T}(z)$ for the up and down quark is around several percent, and the sign is negative (note that $D_1(z)$ in the left panel has been sign reversed); while $D^{\perp(1/2)}_{1T}(z)$ for the s quark is consistent with zero.
This is very different from the cases of the unpolarized $\Lambda$ fragmentation function and longitudinally polarized $\Lambda$ fragmentation function, where strange quark content is significant (for $D_1^{\Lambda}$) or dominant (for $G_1^\Lambda$).
The reason for this discrepancy is due to the dominance of the axial-vector diquark contribution to $D^{\perp}_{1T}$ over the scalar diquark contribution in our model. Moreover, $D^{\perp}_{1T}$ for the strange quark receives contribution only from the scalar diquark, as shown in Eq.~(\ref{relation2}).
Another observation is that our calculated $D^{\perp}_{1T}$ for the up and down quarks does not always satisfy the positivity bound, i.e, at the large $z$ region ($z>0.75$) the bound is violated.
We note that similar violations of the positivity bound were also observed in Refs.~\cite{Pasquini:2014ppa,Wang:2017onm}.
An explanation was given in Ref.~\cite{Pasquini:2011tk}, stating that the violation may arise from the fact that T-odd TMD distributions or fragmentation functions are evaluated to $\mathcal{O}(\alpha_s)$, while in model calculations T-even TMD functions are usually truncated at the lowest order.

In the following, we apply the model result for $D_{1T}^\perp$ of the $\Lambda$ hyperon to predict the transverse $\Lambda$ polarization $P_T^\Lambda$ in SIDIS and SIA, which is a direct experimental observable.
Usually $P_T^\Lambda$ in high energy processes is defined as
 \begin{align}
 P^\Lambda_T={d\Delta\sigma\over d\sigma}={[d\sigma(S_{\Lambda\,T})-d\sigma(-S_{\Lambda\,T})]\over [d\sigma(S_{\Lambda\,T})+d\sigma(-S_{\Lambda\,T})]}\,.
 \end{align}

If only the transverse momentum of the fragmenting quark is considered, one can give a simplified expression for $P_T^\Lambda$ in SIDIS~\cite{Anselmino:2001js}
\begin{align}
P^\Lambda_T(x,y,z,P_T)\big{|}_{\textrm{DIS}}&={\sum_q\,e_q^2\,f_{q/p}(x)[d\sigma^{\ell\,q}/dy]\Delta D_{\Lambda^\uparrow/q}(z,\bm{P}_T^2)\over \sum_q\,e_q^2\, f_{q/p}(x)[d\sigma^{\ell\,q}/dy]D^{q\rightarrow\Lambda}_{1}(z,\bm{P}_T^2) }\,,
\end{align}
where $f_{q/p}(x)$ is the usual unpolarized distribution function, and $d\sigma^{\ell\,q}/dy$ is the lowest order partonic cross section.

\begin{figure}
  \centering
  \includegraphics[width=0.48\columnwidth]{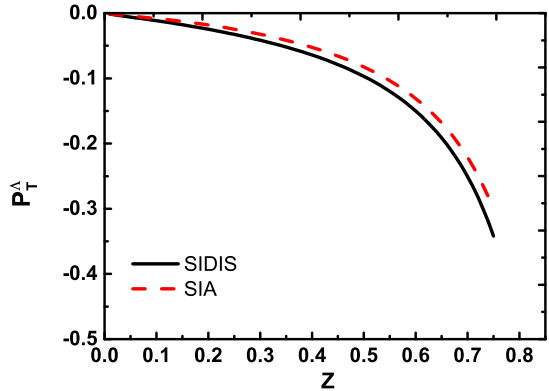}
  \caption{The transverse $\Lambda$ polarizations, $P_T^\Lambda$ vs $z$, averaged over $P_{\Lambda\,T}$, for SIDIS (solid line) and SIA (dashed line).}
  \label{PT}
\end{figure}

Similarly, the transverse polarization $P_T^\Lambda$ in SIA can be written as
\begin{align}
P^\Lambda_T(y,z,P_T)\big{|}_{\textrm{SIA}}={\sum_q\,e_q^2\,[d\sigma^{e^+e^-}/dy]\Delta D_{\Lambda^\uparrow/q}(z,\bm{P}_T^2)\over \sum_q\,e_q^2\,[d\sigma^{e^+e^-}/dy]D^{q\rightarrow\Lambda}_{1}(z,\bm{P}_T^2)}\,.
\end{align}

Using the $SU(3)_f$ symmetric unpolarized $\Lambda$ fragmentation functions, i.e., $D^{{u\rightarrow\Lambda}}_{1}\equiv D^{d\rightarrow\Lambda}_{1}\equiv D^{s\rightarrow\Lambda}_{1}$,
and ignoring the sea quarks $f_{\bar{q}/p}$ and strange quark $f_{s/p}$ contributions to SIDIS,
we give an approximate result for $P_T^\Lambda$ in SIDIS:
\begin{align}
P^\Lambda_T\big{|}_{\textrm{DIS}} \approx {\Delta D_{\Lambda^\uparrow/\textrm{u}}\over D^{\textrm{u}\rightarrow\Lambda}_{1}}. \label{eq:sidis}
\end{align}

Similarly, $P_T^\Lambda$ in SIA has the simplified form:
\begin{align}
P^\Lambda_T\big{|}_{\textrm{SIA}}= {5\Delta D_{\Lambda^\uparrow/\textrm{u}}+ \Delta D_{\Lambda^\uparrow/\textrm{s}}\over 5D^{\textrm{u}\rightarrow\Lambda}_{1}+D^{\textrm{s}\rightarrow\Lambda}_{1}}. \label{eq:sia}
\end{align}

In Fig.~\ref{PT} we plot the transverse $\Lambda$ polarization $P_T^\Lambda$ vs $z$, in both SIDIS (solid line) and SIA (dashed line), after averaging over the transverse momentum of the $\Lambda$ hyperon.
We present the result in the region $z<0.75$, where $D_{1T}^{\perp}$ does not violate the positivity bound. The numerical results show that the transverse polarization of $\Lambda$ is negative in both SIDIS and SIA, and
in both cases the size of $P_T^\Lambda$ increases with increasing $z$.  In the large $z$ region, $P_T^\Lambda$ is substantial.
Our results are consistent with the phenomenological analysis presented in Ref.~\cite{Anselmino:2001js} and with the calculation of Ref.~\cite{Boer:2010ya}.
Furthermore, the shape of $P_T^\Lambda$ in SIA is very similar to that in SIDIS. This is a consequence of the up and down quark dominance for $D_{1T}^{\perp}$ in our model.
The difference between $P_T^\Lambda$ in SIA and SIDIS is given by $D_{\Lambda^\uparrow/\textrm{s}}$, as can be seen from Eqs.~(\ref{eq:sidis}) and (\ref{eq:sia}).
Thus this difference may provide a test for the strange quark contribution to $P_T^\Lambda$.

\section{Discussion and Conclusion}

In this work, we studied the T-odd transversely polarized fragmentation function $D^\perp_{1T}$ for the process $q\rightarrow \Lambda^\uparrow+X$.
We performed the calculation in the diquark spectator model, and we used the relation between the quark flavors and diquark types for fragmentation functions, motivated by the SU(6) symmetric wavefunctions of the $\Lambda$ hyperon.
We obtained the values of the model parameters, by fitting the resulting $D_1^\Lambda(z)$ to the DSV parametrization for $D^\Lambda_1$ at the initial scale $\mu_0^2=0.23$ GeV$^2$.
Using the same model and these parameters, we computed $D^\perp_{1T}$ of the $\Lambda$ hyperon for light flavors.
As a byproduct, we also calculated the longitudinally polarized fragmentation function $G^\Lambda_1(z)$.
The flavor dependence of the different fragmentation functions is quite different in our model. In the case of the unpolarized fragmentation function, $D_1^{\textrm{u}\to\Lambda}(z)$, $D_1^{\textrm{d}\to\Lambda}(z)$ and $D_1^{\textrm{s}\to\Lambda}(z)$, they turn out to be the same.
The longitudinally polarized fragmentation function, $G^{\textrm{s}\to\Lambda}_1(z)$, is positive and sizable, and $G^{\Lambda}_1(z)$ vanishes for up or down quarks, which
is consistent with scenario 1 set of the DSV parametrization for $G^\Lambda_1(z)$.
The situation is opposite in the case of $D^\perp_{1T}$, for which the up or down quark dominate over the strange quark.
Using our numerical result of $D^\perp_{1T}$,  we estimated the transverse polarization $P_T^\Lambda$ in both SIDIS and SIA, and found that in these two processes the polarizations are negative and substantial in the large $z$ region.

Finally, some comments about our calculation are in order. Firstly, in our model the flavor dependence of the fragmentation functions was obtained based on the assumption of SU(6) symmetry of the octet baryons;
secondly, in the calculation of $P_T^\Lambda$, we only considered the leading-order result, and we assumed the same evolution for $D_{1T}^{\perp}$ and $D_1$.
We note that SU(6) symmetry breaking,  higher order corrections and evolution effects for $D_{1T}^{\perp}$ may alter the results only quantitatively, but we fully expect that they will not change qualitatively.

\section{Acknowledgements}

This work is partially supported by the NSFC (China) grant 11575043, by the Fundamental Research Funds for the Central Universities of China, by Fondecyt (Chile) grants 1140390 and FB-0821.

\end{document}